\documentclass{article}
\usepackage{spconf,amsmath,graphicx}
\usepackage{amsfonts,amssymb}
\usepackage{booktabs}
\usepackage{bbm}
\usepackage[colorlinks,
linkcolor=blue,
anchorcolor=green,
citecolor=green]{hyperref}

\title{Personalized speech enhancement combining band-split RNN and speaker attentive module}
\usepackage{spconf}
\makeatletter
\def\@name{ \emph{Xiaohuai Le$^1$$^,$$^2$, Li Chen$^2$, Chao He$^2$, Yiqing Guo$^2$, Cheng Chen$^2$, Xianjun Xia$^2$, Jing Lu$^1$$^,$$^3$} }
\makeatother
\address{
	$^1$Key Laboratory of Modern Acoustics, Nanjing University, Nanjing 210093, China\\
	$^2$RTC Lab, ByteDance, China \\
	$^3$NJU-Horizon Intelligent Audio Lab, Horizon Robotics, Beijing 100094, China}
\begin{document}
\ninept
\maketitle

\begin{abstract}
	Target speaker information can be utilized in speech enhancement (SE) models to more effectively extract the desired speech. Previous works introduce the speaker embedding into speech enhancement models by means of concatenation or affine transformation. In this paper, we propose a speaker attentive module to calculate the attention scores between the speaker embedding and the intermediate features, which are used to rescale the features. By merging this module in the state-of-the-art SE model, we construct the personalized SE model for ICASSP Signal Processing Grand Challenge: DNS Challenge 5 (2023). Our system achieves a final score of 0.529 on the blind test set of track1 and 0.549 on track2.
\end{abstract}
\begin{keywords}
	Personalized Speech Enhancement, Auditory perceptual attention, BSRNN
\end{keywords}
\vspace{-0.2cm}
\section{Introduction}
\label{sec:intro}
The aim of personalized speech enhancement (pSE) is to suppress background noises, reverberation and speaker interference for better desired speech extraction. ICASSP Signal Processing Grand Challenge: DNS Challenge 5 (2023)\footnote{https://aka.ms/5th-dns-challenge} is organized to promote research and development in the field of SE and pSE\footnote{This work was supported by the National Natural Science Foundation of China (Grant No. 12274221).}.

Different from speaker-independent SE, pSE requires additional target enrollment speech. Recent works like TEA-PSE \cite{ref1} are based on combining the speaker embedding, usually obtained by a well-trained speaker verification (SV) model, with effective SE models. Apart from improving the robustness of the speaker embedding, it is also necessary to improve the interaction between the embedding and the SE model for better pSE performance. Concatenating the embedding with the intermediate features is a commonly used method, but the performance is limited due to the distribution mismatch. Affine transformation as used in pVAD \cite{ref3} is expected to have better performance, but it still cannot completely solve the mismatch problem. 

Inspired by Deep attractor network \cite{ref4}, which mimics the auditory attention mechanism and clusters the representations of different speakers to get the target masks, we propose a speaker attentive module (SAM). The module calculates the attention scores between the speaker embedding and different sub-band representations of the features. The scores are then used to rescale the features. Combined with the state-of-the-art (SOTA) real-time SE model, Band-split RNN (BSRNN) \cite{ref5}, we build a pSE system which ranks top 5 in DNS challenge 5 .

\section{Models}
\label{sec:format}

\subsection{Band-split RNN}

\begin{figure}[tbh!]
	\centering
	\vspace{-0.5cm}
	\includegraphics[width=0.5\textwidth]{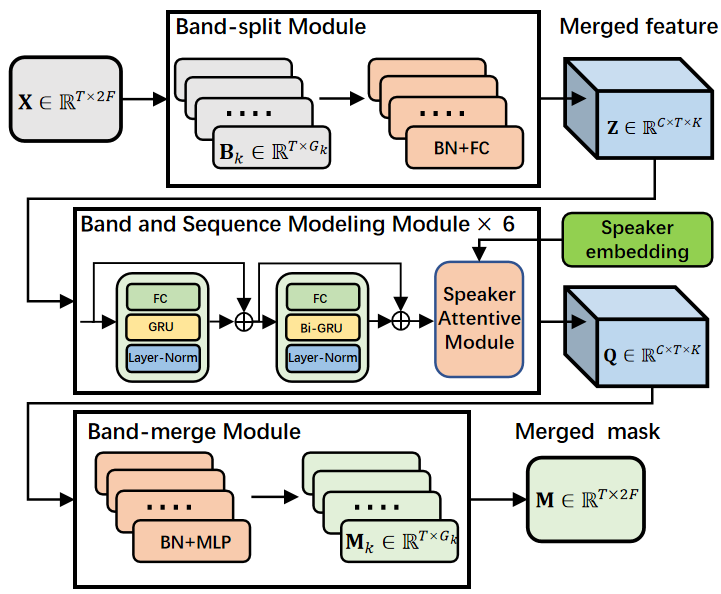}
	\caption{The diagram of the proposed pBSRNN model.}
	\label{fig:figure1}
    \vspace{-0.5cm}
\end{figure}

BSRNN is a SOTA model in real-time SE \cite{ref5}, which consists of the band-split module, band and sequence modeling module, and the band-merge module. As shown in Fig.~\ref{fig:figure1}, we keep the backbone of the BSRNN in our system. The band-split module splits the input noisy spectrogram into \(K\) sub-bands at non-linear intervals, which are later processed by batch normalization layers and linear layers. The sub-band feature is spliced into a 3-dimensional tensor, which is fed into the band and sequence modeling module. The band and sequence modeling module stacks 6 DPRNNs to alternately model the inter-frame and inter-band features. The estimation masks are obtained through a band-merge module symmetrical to the band-split module. In our proposed pBSRNN, ECAPA-TDNN \cite{ref6} is applied to obtain the speaker embedding, and the speaker attentive module is added after each DPRNN.

\subsection{Speaker attentive module}
The core idea of the speaker attentive module (SAM) is to use the speaker embedding as the attractor, and calculate its connection with all sub-band features across the F-dimension. The obtained attention scores are used to rescale the features. Suppose that \(\mathbf{h}\in \mathbb{R}^{B{\times}C{\times}T{\times}K}\) is the intermediate feature input of the SAM and \(\mathbf{e}\in \mathbb{R}^{B{\times}C_2}\) is the target speaker embedding with \(B\), \(C\), \(T\) and \(K\) represent the batch size, the channel number, the frame number and the band number, respectively. The channel numbers of \(\mathbf{h}\) and \(\mathbf{e}\) are first transformed to \(C_1\) to obtain the key \(\mathbf{k}\) and the query \(\mathbf{q}\) as
\begin{equation}
	\mathbf{k}=Expand(FC(\mathbf{e}))\in \mathbb{R}^{B\times{T}\times{C_1}\times{1}} \label{eq1}
\end{equation}
\begin{equation}
	\mathbf{q}=Permute(Conv_0(\mathbf{h}))\in \mathbb{R}^{B\times{T}\times{K}\times{C_1}} ,\label{eq2}
\end{equation}
where \(FC(\cdot)\) denotes the fully-connected layer and \(Conv_0(\cdot)\) is a 2-D convolutional layer with the causal padding, which helps to capture more contextual information. Note that the time frame number of \(\mathbf{k}\) are expanded to \(T\) 
to match the dimension of \(\mathbf{q}\), so that they can be multiplied along the channel dimension to get the attention scores as
\vspace{-0.2cm}
\begin{equation}
	\mathbf{s} = Softmax(\frac{{\sum_c{\mathbf{q}[:,:,:,c]\mathbf{k}[:,:,c,:]}}}{\sqrt{C_1K/2}}) \in \mathbb{R}^{B\times{T}\times{K}\times{1}} \label{eq3}
\end{equation}
Then the attention scores \(\mathbf{s}\) are expanded along the channel dimension to rescale \(\mathbf{h}\), followed by a point convolutional layer with the skip-connection to get the final output \(\mathbf{h}_o\in \mathbb{R}^{B\times{T}\times{K}\times{C}}\) as
\begin{equation}
	\mathbf{h}_o=Conv_1(\mathbf{s}\odot{\mathbf{h}})+\mathbf{h} \label{eq4}
\end{equation}
where \(\odot\) represents the element-wise multiplication. Batch normalization and PReLU are applied in all the convolutional layers in the SAM. We use depth-separable convolution to build \(Conv_0(\cdot)\), leading to a lower computational cost.
\vspace{-0.2cm}
\subsection{Loss function}
The weighted sum of the asymmetric magnitude MSE (\(MSE_a(\cdot,\cdot)\)) \cite{ref1} and the complex compressed MSE (\(MSE_c(\cdot,\cdot)\)) \cite{ref7} is the loss function, denoted as
\vspace{-0.2cm}
\begin{equation}
	L_{se}= 0.3 \times MSE_a\left(\left|S\right|^c,\left|{\hat{S}}\right|^c\right) + 0.7 \times MSE_c\left(S^c,\hat{S}^c\right)) \label{eq5}
 \vspace{-0.2cm}
\end{equation}
where \(S\) and \(\hat{S}\) denote the clean and the output spectrogram, \(S^c\) and \({\hat{S}}^c\) denote the complex spectrogram after magnitude compression.
\vspace{-0.3cm}
\section{Experiments}
\label{sec:pagestyle}
\subsection{Datasets and settings}
We train our model on DNS-5\footnote{https://github.com/microsoft/DNS-Challenge} dataset and Didispeech \cite{ref6}. We note that some of the speech signals are actually corrupted, so we apply a pre-trained SE model on these signals before training. During the training stage, we generate 100,000 4-second noisy reverberated speech clips on-the-fly. The mixing proportions of noise, interfering speakers and target speakers are consistent with those of TEA-PSE \cite{ref1}. The speech with the early reverberation is used as the target. The enrollment speech is randomly cut from the sample of the target speaker and is fed into a pre-trained ECAPA-TDNN\footnote{https://huggingface.co/speechbrain/spkrec-ecapa-voxceleb} to get the speaker embedding. One-tenth of the generated audio will be used for validation. 

The frame length of STFT is 20 ms, and the hop length 10 ms. We utilize the best band split bandwidths in \cite{ref5} and \(K=41\). The channel number of the embeddings is 192. The hidden size of the GRU and the output MLP are 128 and 512, respectively. The model is trained by the Adam optimizer with an initial learning rate of 1e-3. The learning rate halves if the validation loss epochs no longer improve for 10 consecutive epochs.
\vspace{-0.2cm}
\subsection{Results}
\begin{table}[!htbp]
	\vspace{-0.5cm}
	\setlength{\abovecaptionskip}{2pt}
	\setlength\tabcolsep{5pt}
	\caption{The results on the blind test set of both tracks.}
	\label{tab:table1}
	\centering
	\begin{tabular}{ccccccccl}
		\toprule
		Tracks   & Method      & SIG      & BAK   & OVRL  & WAcc & M             \\
		\midrule
		           &  noisy        & 3.76   & 1.22  & 1.22   & 0.843    & 0.449  \\
		Track1    &  Baseline\_p  & 3.20     &  2.67  &   2.34   &   0.687   & 0.511   \\
		           &  Ours         & 3.51     &  2.48  &   2.30   &   0.733   & 0.529  \\
		\midrule
	      		  &  noisy        &  3.83   &  1.22  & 1.24   & 0.857    & 0.459 \\
		Track2      &  Baseline\_p  & 3.22     &  2.68  &    2.38  &   0.727   & 0.537  \\
		   		  &  Ours         & 3.53     &  2.58  &    2.39  &   0.749   & 0.549 \\
		\bottomrule
	\end{tabular}
\vspace{-0.5cm}
\end{table}

The results on the blind test set of both tracks are presented in Table~\ref{tab:table1}.  We submitted results from the same model in both tracks, where Track2 has a higher proportion of near-field speech, leading to higher performance. Compared with noisy speech, it can be seen that our model effectively attenuates the noise and interference and obtains significantly higher scores on BAK and OVRL. Compared to the official baseline pSE model (Baseline\_p), our model achieves a slightly lower score in BAK, but a higher score in SIG. This is advantageous for automatic speech recognition (ASR) and results in a comparatively higher WAcc and the final score M. One possible reason for the lower BAK score is the relatively high weight of the asymmetric MSE. According to our subjective test, it appears that our model is not sufficiently powerful to effectively suppress non-overlapping interfering speakers in longer audio clips. This may be attributed to the lack of this scenario in the training data. Our model finally ranks 4-th in Track1 and 5-th in Track2. The trainable parameters and the computational complexity of the proposed model is about 5.97 M and 5.54 G MACs. The RTF of the exported ONNX model is 0.41 on a single thread of an Intel Core-i5 CPU clocked at 2.4G Hz.

\vspace{-0.4cm}
\section{Conclusions}
\label{sec:typestyle}
In this paper, we propose a real-time pSE model based on the combination of BSRNN and the speaker attentive module, which is used to rescale the subband features in the BSRNN through the attention mechanism. Results on the DNS-5 datasets validate the advantage of our model in signal quality over the baseline. More ablation experiments on the speaker attentive module will be presented in the future work.
\vspace{-0.5cm}

\bibliographystyle{IEEEbib}
\bibliography{refs_dns}

\end{document}